\documentclass{aa}
\usepackage{graphicx}
\usepackage{txfonts}
\usepackage{natbib}
\usepackage{longtable}

\begin{document}

\def\logg{$\log (g)$~}
\def\Teff{$T_{\rm eff}$~}
\def\FeH{$[Fe/H]$~}
\def\vmicc{$\xi_{t}$}
\def\vmic{$\xi_{t}$~}
\def\msun{$M_{\odot}$~}

\def\gtsim {>\kern-1.2em\lower1.1ex\hbox{$\sim$}~}   
\def\ltsim {<\kern-1.2em\lower1.1ex\hbox{$\sim$}~}   


\title{Manganese spread in Ursa Minor as a \\proof of sub-classes of
  type Ia supernovae}

\author {G. Cescutti\inst{1,3} \thanks {email to: cescutti@oats.inaf.it} 
\and  C. Kobayashi\inst{2,3} }
\institute{
INAF, Osservatorio Astronomico di Trieste, I-34131 Trieste, Italy
\and Centre for Astrophysics Research, School of Physics, Astronomy and
Mathematics, University of Hertfordshire, College Lane, Hatfield AL10
9AB, UK
\and BRIDGCE UK Network (www.bridgce.net), UK
}

\date{Received xxxx / Accepted xxxx}

\abstract{Recently, new sub-classes of Type Ia supernovae (SNe Ia) were
  discovered, including SNe Iax. The suggested progenitors of SNe Iax
  are relatively massive, possibly hybrid C+O+Ne white dwarfs, which
  can cause white dwarf winds at low metallicities. There is another
  class that can potentially occur at low or zero metallicities;
  sub-Chandrasekhar mass explosions in single and/or double degenerate
  systems of standard C+O white dwarfs. These explosions have
  different nucleosynthesis yields compared to the normal,
  Chandrasekhar mass explosions.}  {We test these SN Ia channels using
  their characteristic chemical signatures.}  {The two sub-classes of
  SNe Ia are expected to be rarer than normal SNe Ia and do not affect
  the chemical evolution in the solar neighbourhood; however, because
  of the shorter delay time and/or weaker metallicity dependence, they
  could influence the evolution of metal-poor systems. Therefore, we
  have included both in our stochastic chemical evolution model for
  the dwarf spheroidal galaxy Ursa Minor.}{The model predicts a
  butterfly-shape spread in [Mn/Fe] in the interstellar medium at low
  metallicity and - at the same time - a decrease of [$\alpha$/Fe]
  ratios at lower [Fe/H] than in the solar neighbourhood, both of
  which are consistent with the observed abundances in stars of Ursa
  Minor.}{The surprising agreement between our models and available
  observations provides a strong indication of the origins of these
  new sub-classes of SNe Ia. This outcome requires confirmation by future
  abundance measurements of manganese in stars of other satellite
  galaxies of our Milky Way. It will be vital for this project to
  measure not the most extreme metal-poor tail, as more commonly
  happens, but the opposite; the metal-rich end of dwarf spheroidals.}


\titlerunning{Mn spread in Ursa Minor as proof of sub-classes of type Ia SNe}

\authorrunning{Cescutti and Kobayashi}

\maketitle

\section{Introduction}

The scenario leading to a supernova Type Ia (SN Ia) explosion is still
under debate. The two most common progenitor scenarios are: the single
degenerate scenario, in which a white dwarf (WD) with a mass close to
the Chandrasekhar (Ch) limit accretes mass from a companion (a red
giant or a main sequence star), and the double degenerate scenario in
which two WDs merge due to the loss of angular momentum.  Both
scenarios present pros and cons, but at present neither is able to
explain all observational constraints of SNe Ia
\citep[e.g.][]{Maoz14}.  Thanks to observational surveys, a large
number of SNe Ia are observed as a luminous and (almost) standard
candle, and the SN Ia explosions were fundamental to understanding the
expansion rate of the Universe and led to the discovery of dark
energy \citep{Riess, Schmidt,Perlmutter}, after the dismissal by Albert Einstein several decades ago.
On the other hand, the surveys have revealed a small variation of the
standard candle such as `super-luminous` or `faint', referred to as
Type Iax \citep[e.g.][]{Foley13}.

  In this paper, we test possible channels in which different
  scenarios for SNe Ia are actually real and present themselves in the Universe (or
  at least in the chemical pollution).  Theoretically, SN Ia channels
  consist of the progenitors and the explosion mechanism, and the
  former determines the lifetime, while the latter determines the
  nucleosynthesis yields. Observationally, the majority of
  SNe Ia seem to  be Ch-mass explosions \citep[e.g.][]{Scalzo14},
such as deflagrations or delayed detonations \citep{Iwamoto99}.
Although both single and double
  degenerate binary systems have been debated as possible progenitors of SNe Ia,
numerical simulations of
  double-degenerate systems show the explosions to be similar to sub-Ch
  mass detonations \citep{Pakmor12}. 

The class of SNe Iax, also called 02cx-like, is used for various
  objects typically $\gtsim 1$ mag fainter than normal SNe Ia;
  although SN 2002cx is the prototype \citep{Li2003}, this class
  includes objects ranging from $M_{V} \sim -19$ to $-14$ mag at peak
  \citep[e.g.][]{Miller17}.  There is one detection of the companion
  star \citep{McCully14}, and the host galaxies tend to be late type
  \citep{Foley13,White15}.  Although several
    models have been discussed for particular objects, the preferred
    model for this class is a single degenerate system of a hybrid,
    C+O+Ne WD \cite[e.g.][]{Meng14}.  In the stellar
    evolution of intermediate-mass stars, such hybrid WDs are known,
    and the outer O+Ne+Mg layer gives a natural reason for the small
    mass of ejecta with an undisrupted WD after the explosion.  The
    initial mass range of hybrid WDs depends on the metallicity, as
    well as the mass-loss and nuclear reaction rates.  
     In any case, the rate is expected to be higher for lower
      metallicities with hybrid WDs.
\citet{Kobayashi15}
    first included SNe Iax in galactic chemical evolution models,
and we use the same mass ranges as adopted in that publication
for the primary and secondary stars in this paper.
  Since the WD mass is already close to the Ch-mass, WD
    winds can easily occur, and thus a wide range of secondary mass is
    allowed \citep[Fig. 2 of][]{Kobayashi15}.
    Sub-Ch mass C+O WDs ($\sim 1M_\odot$) have been invoked to
      explain some spectral features of SNe Ia
      \citep[e.g.][]{Shigeyama92} and also discussed in relation with
      the rate of SNe Ia predicted by binary population synthesis
      \citep{Ruiter14}. 
      Both single and double degenerate systems are possible. The
      carbon detonation is triggered by detonation of a thin He layer,
      and the ejected iron mass is not so different from normal SNe
      Ia.  In the single degenerate systems, the required mass
      accretion rate is lower than Ch-mass WDs, where there is no WD
      wind, and the rate does not have a strong dependency on the metallicity
     \citep[Fig. 1 of][]{Kobayashi15}.
 Therefore, these two channels
    (SNe Iax and sub-Ch SNe Ia) are dominant at low metallicities,
    which is in contrast to normal SNe Ia (with Ch-mass C+O WDs in
    this paper)  that
    are inhibited due to the lack of WD winds at [Fe/H] $< -$1.1
    \citep{Kobayashi98,Kobayashi09}.

 In general, it is not easy to constrain SN Ia scenarios because the
 chemical outcomes are similar.  However, the chemical signature
 is relatively clear in the case of Mn, which is one of the few
chemical elements with just a single stable isotope ($^{55}$Mn).
Recently, this chemical element has been investigated in
\citet{Kobayashi15}, where the presence of 
  three channels (Ch, sub-Ch, and Iax) for the production have been shown to promote a different trend in
the chemical evolution for dwarf galaxies, keeping the trend
in the case of the solar vicinity model unchanged. A similar analyses was
performed, but only for the solar vicinity, in \citet{Seitenzahl13},
and the final results were in favor of the presence of two 
scenarios for SNe Ia (Ch and sub-Ch) . Regarding the results for  dwarf galaxies, a
flat and under solar trend for Sagittarius was obtained by
\citet{Cesc08b}, by means of a strong dependence to the metallicity of
the SN Ia yields, maintaining again a good agreement for the solar
neighbourhood. The same observational results were obtained for
Sculptor, Fornax, Carina and Sextans in \citet{North12}.   We note
  that, however, such a metallicity effect is not expected for the
  majority of SNe Ia where Mn is synthesised in nuclear statistical
  equilibrium \citep{Kobayashi06,Kobayashi15}.

  The final results are indeed connected to the assumptions regarding
  the rates of the different channels which are indeed still quite
  difficult to be constrained from observational data and theoretical
  one-zone models \citep[e.g.][]{Matteucci01}.  However, in the present
work we intend to deal with another possible aspect by using
a stochastic model:
the spread in the [Mn/Fe] ratios.
We will show both types of model in Section 5.

Previous theoretical works were able to use the spread observed in the
Galactic halo to study the production of chemical elements, in
particular neutron capture elements \citep{Tsujimoto99,
  Ishimaru99,Trava01,Argast02,Argast04,KaGu05,Cesc08a,
  Cesc10,Cescutti13}.  Indeed  stochastic modelling enables us to gain an
approximative but quantitative description of the spread produced by
the different sources polluting the inter stellar medium (ISM). This
is particularly useful for rare sources, and the stochastic modelling
of the Galactic halo was able to reproduce, for example, the spread in
neutron capture elements - due to the rare r-process events \citep[see
also][]{Cescutti14,Cescutti15} - and the smaller dispersion in the
alpha-elements regularly injected into the ISM by SN II explosions at
the same time. The impact of rotating
massive stars on the chemical evolution of CNO \citep{Cesc10} and of faint SNe
on the evolution of carbon and Ba \citep{Cescutti16} has been also
investigated by means of stochastic modelling.
.  We note that, there is the third approach,
  chemodynamical simulations of galaxies \citep[e.g.][]{Kobayashi11b}, 
in which hydrodynamics determine the star formation histories
  and chemical enrichment takes place inhomogeneously. Since these
  simulations take a lot of computational time, it is not yet
  possible to explore all possible sources of chemical enrichment.

  In this work we use the stochastic and inhomogeneous chemical
  evolution models to study the pollution due to SNe Ia, not in
  the early stage of the formation of the Galactic halo, but in the
  later stage of a dwarf spheroidal (dSph) galaxy, a satellite of our
  Galaxy.  At present, not many measurements of Mn are present in
  the literature for stars in dSph galaxies.  We decide to compare our
  results with Ursa Minor because it is one of the dSph galaxies
  with the largest number of stars for which Mn is measured
  \citep{Ural15}, and therefore we decide to apply our modelling to this
  case.

A significant effort in this subject has been carried out in \citet{North12},
where the authors measured Mn in stars belonging to four dSph galaxies:
Sculptor, Fornax, Carina and Sextans.  In this work, they presented an impressive amount of Mn data for
Fornax and Sculptor.  Indeed, we
could have started working on one of these two galaxies; however
Sculptor and Fornax are more massive and metal rich
\citep{McConnachie12} compared to Ursa Minor and it is likely that the
signature we are looking for is clearer in fainter and more metal-poor
objects. On the other hand, Sextans and Carina have
stellar masses and absolute luminosities quite similar to those of Ursa
Minor, and we could have developed a model for one of these
systems, for which however the amount of data is lower than for Ursa
Minor.

\section{SN Ia scenarios and nucleosynthesis}\label{Nucleos}

In this paper, we consider the same assumptions for sub-luminous (such
as SNe Iax), sub-Ch, and Ch-mass SNe Ia in terms of nucleosynthesis
and rates of the events as in \citet{Kobayashi15}. For comparison, we
present also results obtained using the SN Ia prescriptions
  adopted in \citet{Cesc08b} and \citet{Spitoni09},
  which are basically the same prescriptions proposed by the influential 
\citet{Matteucci1986}.  We decide to call the former
prescriptions ``Herts'' model, whereas the second is called
``Trieste'' model, from the location of the groups that have developed
each model.

\subsection{Herts model}

One important difference is that in the \citet{Kobayashi15} work,
because of the lack of WD winds \citep{Kobayashi98} there is no
formation of normal SNe Ia at low metallicity ([Fe/H] $\leq -$1.1),
where normal SN Ia channel is defined as the single degenerate
scenario with near Ch-mass delayed detonations or deflagrations.
However, at low metallicity, both sub-Ch SNe Ia and the new channel
called SNe Iax can explode. Both SNe Iax and sub-Ch SNe Ia play
therefore a role in the production of Fe and Mn in dSphs in this
framework, and we have incorporated the table presented in
\citet{Kobayashi15} for metallicity [Fe/H]$<-$1.1, given chemical
evolution of Ursa Minor does not include stars at higher metallicity.

Therefore, we assume that the SN Iax channel produces M(Fe) = 0.193 \msun
and M(Mn) = 3.67$\times 10^{-3}$ \msun  adopting the
nucleosynthesis yields of the N5def model in \citet{Fink14} \citep[see
also][]{Kromer13}. The range of primary star mass is 
6.5\msun $<$M$_{\rm primary}<$ 7.5 \msun, and the secondary can be
either a red giant in the mass range 0.8\msun $<$M$_{\rm primary}<$ 3\msun
or a main sequence star in the mass range 1.6\msun$<$M$_{\rm
  primary}<$ 6.5\msun, depending on the metallicity.
For both  configurations the probability that the systems explode
as SNe Ia is 0.025.

For the sub-Ch channel, we assume the nucleosynthesis yields taken
from the 1.05 \msun model of \citet{Shigeyama92}: M(Fe)=0.5643\msun,
and M(Mn) = 3.246 $\times$ 10$^{-3}$ \msun at the solar metallicity. These
values are really close to the 1.06 \msun model in \citet{Sim10} and
\citet{Seitenzahl13}. In sub-Ch SNe Ia, Mn is mostly synthesised in
incomplete Si-burning, and therefore the Mn yields should depend on
metallicity. We include this effect as M(Mn) $\propto$ Z$^{3}$ 
  as in the calculations of \citet{Seitenzahl15} and \citet{Yamaguchi15}.

\subsection{Trieste model}

The Trieste model is based on the seminal work of
  \citet{Matteucci1986}, which assumes that the sole normal SNe Ia are
  present but at all metallicities; the mass of
  binary systems that can explode as normal SNe Ia are in the range
  3\msun$<$M$_{{\rm binary}}<$ 16\msun, with the maximum mass of the primary
  of 8\msun.  We assume for this model the W7 yields from
  \citet{Iwamoto99}, M(Fe)=0.6 \msun and M(Mn) = 8.87 $\times$ 10$^{-3}$ \msun.
  For Mn yields of SNe Ia we also consider an empirical dependence on
  the metallicity of the yields as considered in \citet{Cesc08b}:

\begin{equation}
Y_{\rm Mn}(z)=Y_{\rm Mn}^{\rm Iwamoto} \left( \frac{z}{z_{\odot}} \right) ^{0.65}
.\end{equation}

In this model  the shortest timescale for the explosion of a SN Ia
is the evolutionary timescale of 8\msun star, which is about 30Myrs.  We note that this timescale is 
similar to the timescale of SNe Iax in the Herts model.

We intend to investigate not only the different channels and the
  different nucleosynthesis yields, but also the fact that the two
channels of SNe Ia operating at low metallicity in the Kobayashi
framework produce different [Mn/Fe] ratios. For this reason, we expect on
average an intermediate value, but with the presence of spread in the
[Mn/Fe]. Conversely, the single scenario presented in the 
Trieste prescriptions will produce either an increase in [Mn/Fe] if we
consider the standard yields, or a flat trend \citep[see for
comparison the results in][]{North12} if we consider a
dependence on the metallicity of the yields of SNe Ia as considered in
\citet{Cesc08b}.
 In both cases with or without the metallicity dependence, no [Mn/Fe] spread in expected.

\subsection{Other model assumptions}

Concerning the nucleosynthesis of the massive stars, we have used in
the Trieste model the yields from \citet{WW95} at solar metallicity,
but considering  the newly produced yields \footnote{
The newly produced yields for the element i are defined as:
\begin{equation}
Y_{\rm new}(i)=M_{\rm ej} \times (X^{\rm nucl}_{\rm fin}(i)-X^{\rm nucl}_{\rm ini}(i))
,\end{equation}
where $M_{\rm ej}$ is the mass ejected by the star, $X^{\rm nucl}_{\rm fin}(i)$
the final abundance of the element `i'  in the  nucleosynthesis
computation  and $X^{\rm nucl}_{\rm ini}(i)$ its
 original abundance again in the  nucleosynthesis
computation.
 The total stellar yields considered in our code for the element `i' are:
\begin{equation}
Y_{\rm tot}(i)=Y_{\rm new}(i)+M_{\rm ej} \times X^{\rm CE}_{\rm ini}(i)
,\end{equation}
where   $X^{\rm CE}_{\rm ini}(i)$ is the chemical abundance of the element `i'
 in the chemical evolution code when the star was born. In this way,
the computed  ISM is not spuriously polluted by the chemical pattern 
 present in the initial composition of the stellar evolution model.}.
 This has been
done to reproduce the previous results obtained in \citet{Cesc08b},
and in particular for Ursa Minor in \citet{Ural15}.  On the other
hand, in the Herts model, we have used the yields with a metallicity
dependency for SN and hypernovae (HN) taken from \citet{Kobayashi11},
considering that half of the stars with $M\ge20$ \msun explode as HNe
as in the model by \citet{Kobayashi06, Kobayashi11,Kobayashi15}; 
  also in this case, we consider  the newly produced material.

Single stars with masses of M $<$ 8\msun, that is, low-intermediate mass
  stars ending their lives as asymptotic giant branch stars, hardly
produce Mn or Fe.  Therefore, we do not discuss their yields in
detail; however, in this mass range the yields by
\citet{vandenHoek97} are considered in the code of both models and
they have only a small impact on the evolution of the metallicity.

In the Trieste model, we consider the initial mass function (IMF) by
\citet{Salpeter55}, that is, a power law:
\begin{equation}
  \phi(m)\propto m^{-x}
,\end{equation}
with a single slope  x=1.35\footnote{ In galactic chemical evolution models, the IMF
 is usually defined as the mass of stars formed in the mass interval (m,
m+dm), instead of the number of stars formed; for this reason, the Salpeter
slope is 1.35, instead of 2.35.} for  0.1 \msun $\le$ m $\le$ 50 \msun.
The Herts model adopts the IMF from \citet{Kroupa08}
which is again a power-law, but with three slopes at different mass ranges: x = 1.3
for 0.5 \msun $\le$ m $\le$ 50 \msun, x =0.3 for 0.08 \msun $\le$ m $\le$ 0.5 \msun and
x = $-$0.7 for 0.01 \msun $\le$ m $\le$ 0.08 \msun.

\begin{table*}
\hskip 2cm
\begin{minipage}{140mm}

  \caption{The main differences in the chemical evolution between
     Herts and Trieste models  are summarised here.
    In the Herts model the Ch-mass single-degenerate channel is not active for [Fe/H]$<-$1.1.}\label{tab_mod}
\centering
\begin{tabular}{||c|c|c|c|c|c|c||}

\hline
\hline
Model & Fig. & normal & SubCh  & SNIax & IMF & SF law \citep{Ural15}\\
          &        & SNIa    & SNIa      &            &       &  $\nu$          \\ 
\hline
Trieste  &  5   &    yes            &    no               &  no     &  Salpeter & 3 $\times$ 10$^{-2}$\,Myr$^{-1}$\\
\hline
Herts   &   6  &    no             &   yes                &  yes    & Kroupa 08  & 7 $\times$ 10$^{-2}$\,Myr$^{-1}$\\

\hline
\hline
\end{tabular}
\end{minipage}

\end{table*}

\section {Chemical evolution models for Ursa Minor}

We start our analysis on the dSph Ursa Minor from the standard chemical
evolution model described in \citet{Ural15}, in particular their 
model C, which takes into account a star formation history based on
the observational constraints by \citet{Carrera02} and infall and winds from the
system able to reproduce the metallicity distribution function (MDF)
obtained by \citet{Kirby11} \citep[for details, see Sect. 3][]{Ural15}.
 In fact, the [Mn/Fe] predicted in the Trieste model with the yields
 from \citet{Cesc08b} are exactly the same as the results 
shown in this paper. 

\begin{figure}[ht!]
\includegraphics[width=85mm]{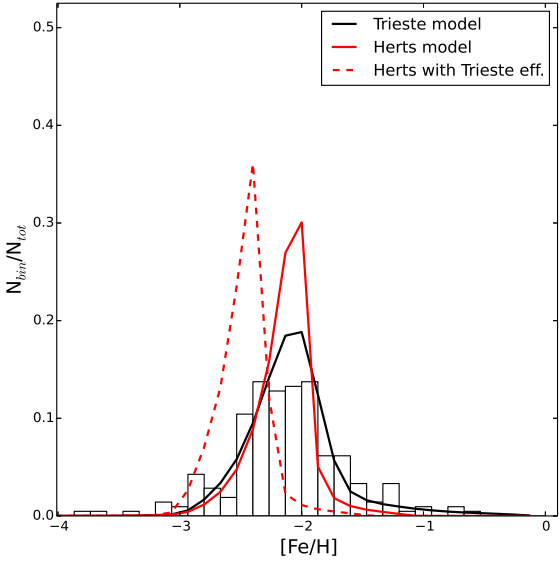}
\caption{Metallicity distribution function (MDF) for Ursa Minor; the histogram shows the 
observational MDF from the data of \citet{Kirby11}. The solid lines are the theoretical
MDFs from Trieste  (black) and Herts (red) models; the dashed line is the MDF for the Herts
model if we keep the star formation efficiency as in the Trieste model.
} \label{MDF}
\end{figure}

Concerning the chemical evolution for the Herts model, we have assumed
a different IMF compared to the Trieste model in order to enable our
modelling to reproduce the results obtained in former work
\citep{Kobayashi15}.  So, in the Trieste model we have a Salpeter IMF
\citep{Salpeter55}, whereas in Herts model we use the
\citet{Kroupa08}. The only other difference between the two models is
the SFR efficiency which is increased by a factor of 2.3 in the Herts
model to reproduce the observational MDF (cf. Fig. \ref{MDF}). This is
because the nucleosynthesis adopted for iron in massive stars is
different in the two models; the iron yields in the Trieste model are
close to the HN set of yields  in the Herts model, but the average
  production of Fe in Herts model is about a factor of two  lower than the
  Trieste model.  For this reason, we have to double the star
formation efficiency.  With these prescriptions both models are able
to reproduce the peak of the MDF and the tail at higher metallicities.
Without this correction, and, therefore, with the same efficiency as the Trieste
model, the Herts model would not produce a MDF compatible with the
observed one, as shown in Fig. \ref{MDF}.  A summary of the
  difference between the Herts and Trieste models is listed in Table
  \ref{tab_mod}.

The target of this paper is however, to investigate the possible
spread produced by a double channel of Mn production from two
different sources.  Therefore, we have developed a stochastic chemical
evolution model in the same fashion as those implemented in
\citet{Cesc08a,Cescutti13} for the Galactic halo, but with the
specific star formation history of Ursa Minor, described above.
This stochastic modelling enables us to gain an approximative but
quantitative description of the spread produced by the different
sources polluting the ISM, and is particularly useful for rare
sources, such as sub classes of SNe Ia (see also Sect. 1). 
As we will see,  in this model we predict not only a dispersion in the 
first enrichment by SNe II, but also later in the chemical evolution due to the
differential production of Mn by the different SNe Ia channels.

\section{Abundances measured in Ursa Minor stars}

We compare our results with the same set of stars shown in
\citet{Ural15}. In this work, the abundances of three stars have been
measured and compared to the abundances collected from other authors
\citep{SCS01,SAI04,Cohen10,Kirby12}. We also decided to compare our final
results with data coming from another two dSphs similar to Ursa
Minor, Sextans and Carina, using the data available in
\citet{North12}. The [Mn/Fe] ratios measured in
  \citet{North12} are shifted by 0.04 dex to account for the different
  solar manganese compared to the one
adopted in \citet{Ural15}.

\section{Results}

\begin{figure}[ht!]
\includegraphics[width=85mm]{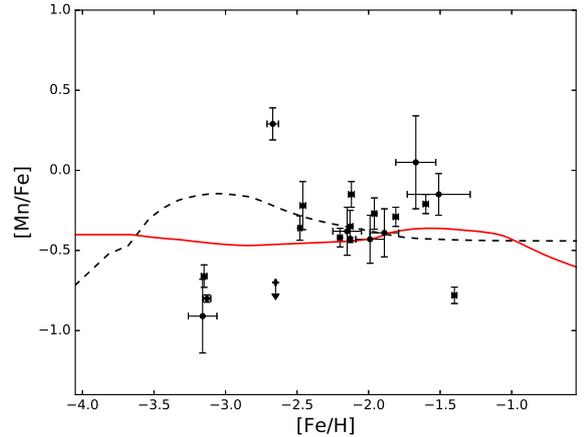}
\caption{ [Mn/Fe] vs. [Fe/H]; the data in black represent the
  abundances measured in stars belonging to the dSph Ursa Minor presented in
  \citet{Ural15}; the red solid line shows the results obtained with the homogeneous
  chemical evolution by the
  Herts model, whereas the dashed black line displays  the results of
  the Trieste model.} \label{fig1}
\end{figure}

\begin{figure}[ht!]
\includegraphics[width=85mm]{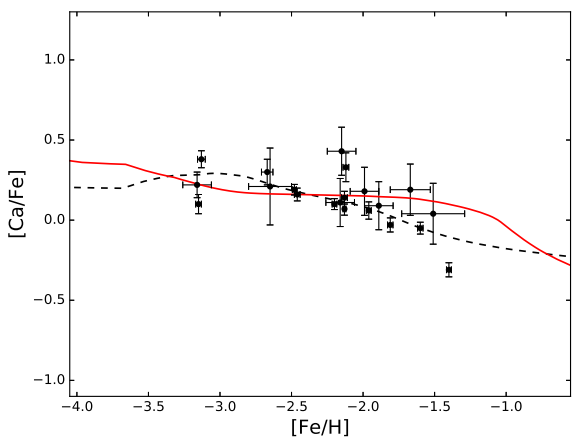}
\caption{[Ca/Fe] vs. [Fe/H]; the data in
  black represent the abundances measured in stars belonging to the
  dSph Ursa Minor presented in \citet{Ural15}. The
  red solid line shows the results obtained with the homogeneous
  chemical evolution by the Herts model, whereas the dashed
  black line displays the results of the Trieste model.} \label{fig2}
\end{figure}

In Figs \ref{fig1} and \ref{fig2}, we present our first results, which are the
comparison between the results of a homogeneous model
of the nucleosynthesis assumed in the Trieste model \citep [as
in][]{Ural15},  and the results of the one assumed in
the Herts model \citep [compatible with][but with a specific model for
Ursa Minor]{Kobayashi15}. 
These Figures do not provide new results, but confirm 
that the Trieste model reproduces the original model
presented in \citet{Ural15} and the Herts model behaves similarly to
the results shown for a generic dSph in \citet{Kobayashi15}.
More importantly, the comparison between the two models shows that in
this homogeneous framework both nucleosynthesis processes are quite compatible
with the data for this dSph.

The low [Mn/Fe], around $-$0.5 dex at [Fe/H]$>-$2.5, 
is caused by different phenomena in the two cases. In the
Trieste model, it is connected to the metal dependency assumed for 
the yields of Mn in SNe Ia. This metal dependency produces 
a small amount of Mn, if the progenitor has a low metallicity (see Eq. 1
),  as is the case for all the progenitor in  Ursa Minor.
In the other case, the Herts model produces  [Mn/Fe] that is relatively flat at a
level of [Mn/Fe]$\sim -$0.4 because it has the two competitive  contributions of 
SNe Ia, producing low [Mn/Fe] and high [Mn/Fe]
\citep[see][, for the models with each contribution]{Kobayashi15}. 

The behaviour of the two models at extremely low metallicity 
is not similar, which is driven by the different nucleosynthesis assumed for massive stars. 
The Trieste model shows relatively large variation, which is due to
the dependency on the stellar mass of Mn ejected by the 
SNe II in the nucleosynthesis adopted for this model \citep[the solar
metallicity yields by][]{WW95}.
In the Herts model, as discussed in Sect.\ref{Nucleos},
 we assume that, on average, half of the massive stars above
20 \msun explode as SNe II and the rest as HNe.
This mix of yields with different energies produces the relatively constant 
[Mn/Fe]; we note, however, that similarly to 
the SN Ia channels, different [Mn/Fe] ratios are produced from SNe II
and HNe, but they are
averaged in the homogenous chemical evolution model.

\begin{figure}[ht!]
\includegraphics[width=95mm]{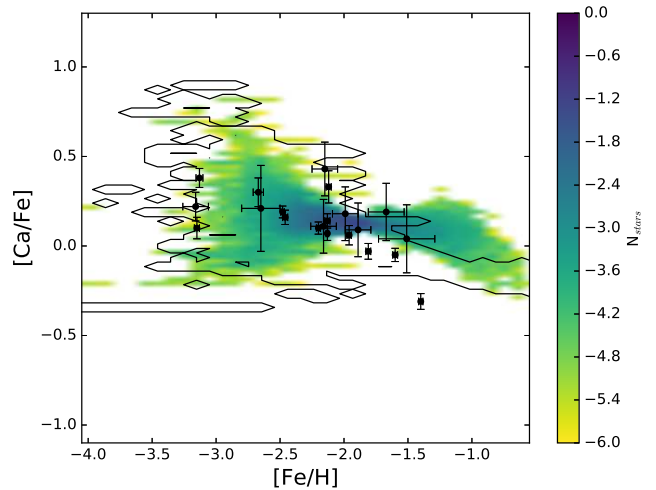}
\caption{[Ca/Fe] vs. [Fe/H]; the data are the same as in
  Fig.\ref{fig2}.
Here we show the results with the stochastic modelling. The
  colour-coded surface density plot presents the density of long-living stars for the Herts model (see colour-bar), whereas the results
  of the Trieste model are displayed with a black
  contour.} \label{fig2b}
\end{figure}

The target of these homogenous models is not to analyse the SNe Ia production of Mn
(and Fe), but to show
that the SN Ia channels are able to reproduce the trend in $\alpha$
elements. After the \citet{Matteucci90}, it is generally accepted that
the delayed SN Ia production is the cause of the knee observed in the
[$\alpha$/Fe] versus [Fe/H] elements in the solar vicinity at around
[Fe/H]$=-$1 and of a similar knee, but at lower metallicities, for
most of the dSph \citep[see][]{Lanfranchi08}.  The same behaviour has
been predicted in the \citet{Kobayashi15}.  Therefore, in Fig.
\ref{fig2}, we show the results of the models for one of the
$\alpha-$elements, calcium. Both models have been slightly shifted
(Trieste 0.1 dex and Herts $-$0.15 dex), to have models pass
through [Ca/Fe]$\sim$ 0.25 at [Fe/H]$\sim -$2.3.  These small offsets are useful
to have a fair comparison between them, assuming 25\%
variation in the yields of Ca can be accommodated in the uncertainties
of the stellar model.  It could also be the signature of a slightly higher
fraction of HN than in the Herts model; in the Trieste model - as mentioned
before - this could be due to the explosion modelling that 
produces too much Fe; indeed dividing the Fe yields by a factor of two has been suggested \citep{Romano10}.

%

 Both models predict a knee for the $\alpha$-elements in
  agreement with the abundances measured in the stars of Ursa Minor. 
In the Trieste model the knee starts early on 
  at [Fe/H] $\sim -$3.0 - although part of this trend could be due to the
massive stars contribution - and continues with a gentle slope up to [Fe/H]
  $\sim -$1 . In the Herts model, it starts only at about [Fe/H]
  $\sim -$1.5 and goes down more quickly. The Trieste model in this
  respect seems to fit the data slightly better than the Herts one.
We recall that in the Herts model at this metallicity there is the contribution
of SNe Iax (producing a low amount of Fe) and 
of sub-Ch SNe Ia, but there is no formation of normal SNe Ia.  

Plots in Figs. 2 and 3 show that both scenarios are able to reproduce the
main observed trends within the uncertainties.  This confirms that - at
least in first approximation - the two models are compatible and it is
not possible to distinguish between them by means of a
 homogenous model

\begin{figure*}[ht!]
\begin{minipage}{180mm}
\includegraphics[width=180mm]{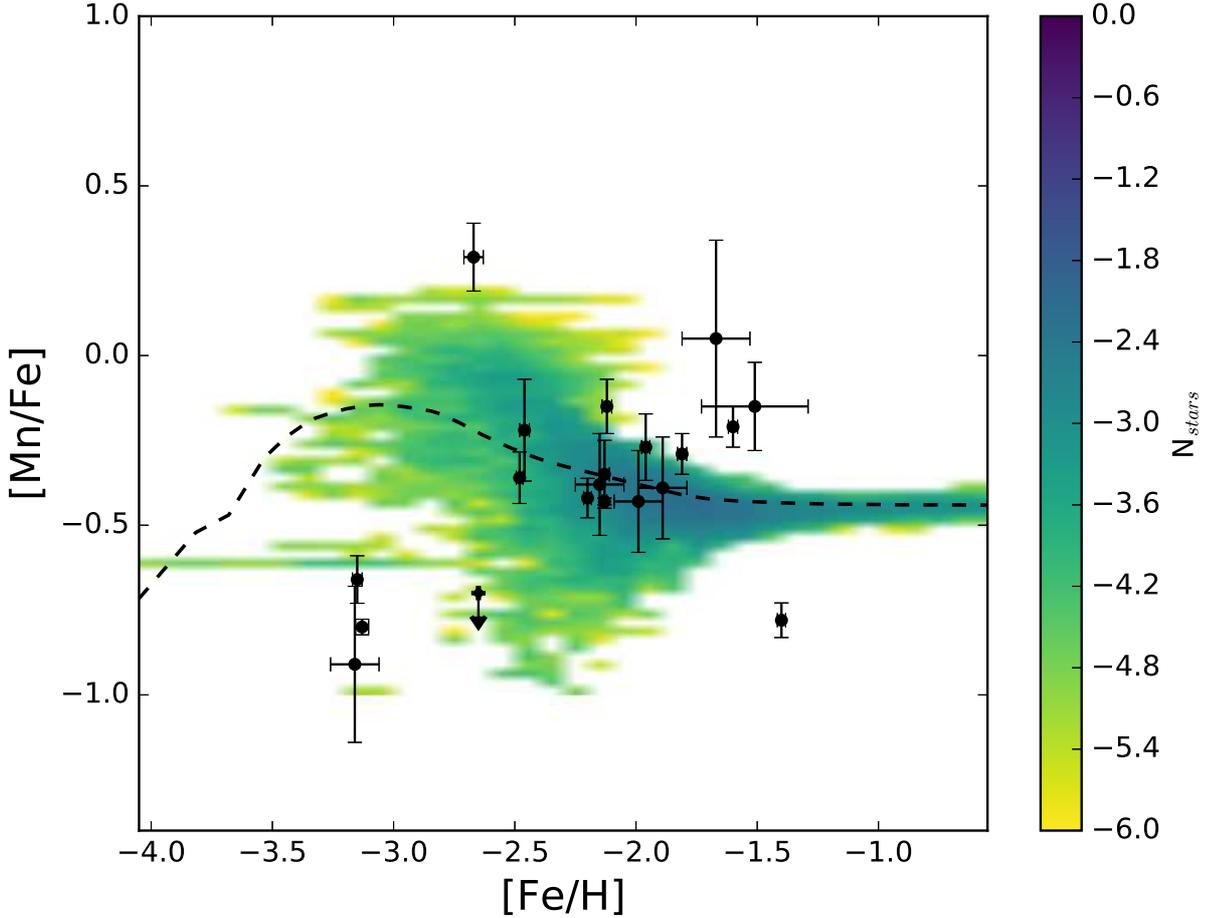}
\caption{[Mn/Fe] vs. [Fe/H]; the data in black represent the
  abundances measured in stars belonging to the dSph Ursa Minor presented in
  \citet{Ural15}. The colour-coded surface density
  plot presents the density of long-living stars for the Trieste model.}\label{fig3}
\end{minipage}
\end{figure*}

%

\begin{figure*}[ht!]
\begin{minipage}{180mm}
\includegraphics[width=180mm]{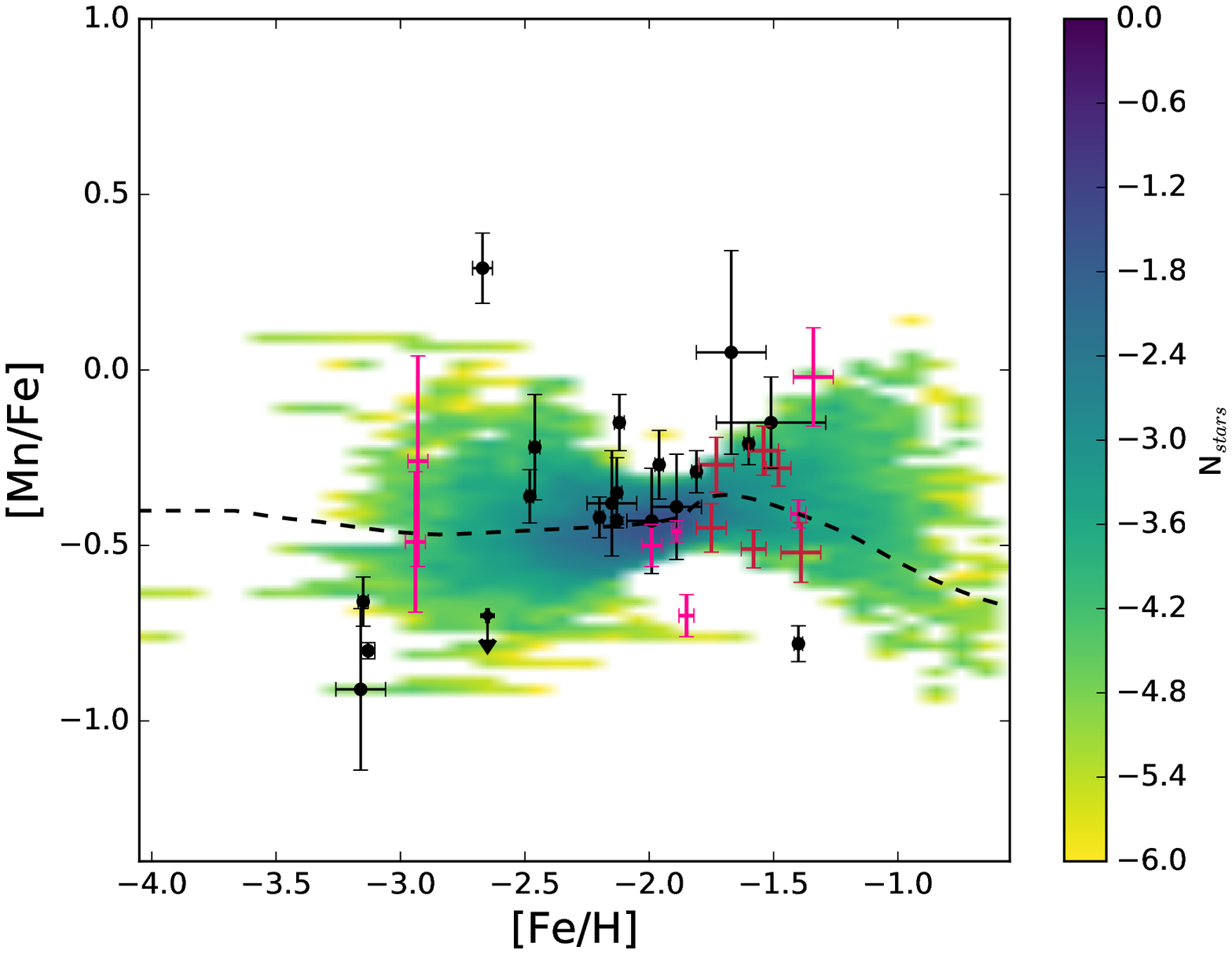}
\caption{ [Mn/Fe] vs. [Fe/H]; the data in black represent the
  abundances measured in stars belonging to the dSph Ursa Minor presented in
  \citet{Ural15}; in red we present the data for the dSph Carina from
  \citet{North12} and
  in magenta those for the dSph Sextans again from from \citet{North12}
  and \citet{TJH10}. The colour-coded surface density
  plot presents the density of long-living stars for the Herts model.
}\label{fig4}
\end{minipage}
\end{figure*}

On the other hand, the use of stochastic modelling could outline a
different conclusion, at least for the Mn case.  In fact, for the
[Ca/Fe] versus [Fe/H] the stochastic model does not produce significant
differences between the two models (see Fig. \ref{fig2b});
both Herts and Trieste models can account for most of the data in Ursa
Minor, with the possible exception of the star with the lowest
[Ca/Fe], which appears an outlier in this plot (and possibly for
the Mn too).

On the other hand, the situation is different in Figs. \ref{fig3} and
\ref{fig4}, where the [Mn/Fe] versus [Fe/H] results for the Trieste and
Herts models using the stochastic approach are presented.

In the Trieste model (Fig. \ref{fig3}), the strong dependence of 
[Mn/Fe] on the stellar mass in massive stars produces a large spread
for [Fe/H]$<-$2.  On the other side, starting from  [Fe/H]$\sim-$2 the
spread is relatively reduced, and this is due to the almost fixed
enrichment in [Mn/Fe] produced by the SNe Ia.  The agreement with most
of the abundances of [Mn/Fe] measured in Ursa Minor - around
[Mn/Fe]$\sim -$0.4 and [Fe/H]$\sim -$2 - is relatively good. The spread
of the Trieste model at extremely low metallicity can also explain the
extremely high value of one star with [Mn/Fe]$>$0 at [Fe/H]$\sim-$2.7
and also - but at the edge of the probability distribution - the three
stars with [Mn/Fe]$<-$0.5 (and the star with just an upper limit).  At
[Fe/H]$>-$2, when the SNe Ia start to play an important role, the
spread for the Trieste model in the [Mn/Fe] versus [Fe/H] space is
decreased, due to the approximately constant enrichment of Mn and Fe
from the SNe Ia. In the region [Fe/H] $> -$1.8, the model does not
agree with any of the abundances measured in the four stars.

The  Herts model with its stochastic
results  shown in Fig.\ref{fig4} displays a butterfly shape
distribution with 
remarkable differences compared to  the Trieste model.

Again the bulk of the data  for Ursa Minor at [Fe/H]$\sim-$2 are
within the prediction of the model; moreover at lower metallicities
most of the measured stars are in good agreement with the model, and
in this case the four stars at [Fe/H]$<-$2.5 and [Mn/Fe]$<-$0.5 are
more clearly within the limits of the probability predicted by the model. In
this region there is, however, a clear outlier: the star with a
[Mn/Fe]$>$0. For such a small sample, excluding one star may not be
ideal. We note that, however,  this star is the only star with a [Mn/Fe]$>$0
at [Fe/H]$<-2.5$ in Ursa Minor, also in other dSphs
\citep{North12}, and possibly the Milky Way; therefore, this
measurement may suffer from some issues. Nevertheless, we consider this
star as a special case and conclude that this model is successful in
this range of metallicity.  The most striking difference is on the
high-metallicity tail. In this region, at [Fe/H]$>-2$, the model
produces again a spread (the right wing of the butterfly). This is
due to the onset of the two SN Ia channels  (SNe Iax and sub-Ch
  SNe Ia) that start to enrich the ISM; we recall that at this
  metallicity, in the Herts model there is no formation of normal SNe
  Ia. Producing  different [Mn/Fe] ratios, they create a spread in
the model results: regions polluted by SNe Iax are extended toward solar
[Mn/Fe] ratios, the contrary for sub-Ch SNe Ia that give low [Mn/Fe].

Comparing the data we have for Ursa Minor, it clearly appears that the
Herts model is the one that more closely approximates what is 
displayed by  the stars of this galaxy. In fact, for [Fe/H] $> -$1.8, three stars 
have a [Mn/Fe]$>-$0.5 and one instead has a lower value,
presenting therefore a spread. At this stage, the amount of data is not yet great enough 
to ensure a statistically sound prediction, and a future observational
campaign to measure more spectra of stars in Ursa Minor is encouraged.
Moreover, we underline that it will be vital for this project to measure
not the most extreme metal-poor tail, as commonly happens, but
the opposite, the metal rich end, in order to disentangle this problem.

Again, we note that the average effect of the two channels
considered in the Herts model are compatible with the average output
of the Trieste model, and therefore only in faint satellites such as Ursa
Minor can this difference be clearly found.
In fact, the expectation is to see this spread being less extreme for galaxies
more massive and with higher average metallicities.
The reason for this expectation comes from two  motivations:
first, the impact of the stochasticity is 
less important at higher metallicities, mixing up the enrichment of
the SNe Ia with more pre-existing Mn and Fe coming from massive stars;
this point can be proven to be wrong depending on the way the
metals are removed by winds (or other mechanism that have acted to
deplete gas from these objects) compared to the timescale of the SN Ia explosions.
The second point is that nucleosynthesis calculations
predict that the amount of Mn produced by the sub-Ch channel
is dependent on the metallicity and therefore at higher metallicity
the difference between the two channels is weaker and spread could be
reduced. 
So, the possible smaller spread in larger dSph galaxies 
could still be investigated as well as in the Galactic
halo; it is necessary to develop specific models to study this cases.

As mentioned in the Introduction, not many measurements of Mn are
present in the literature for stars in dSph galaxies. In this respect a
significant amount of work has been carried out by \citet{North12},
where the measurement of four dSph galaxies are presented: Sculptor,
Fornax, Carina and Sextans.  It appears that also for these galaxies
the spread is present, although specific chemical evolution models
are required to check our theory against these data.  This is
particularly true in the case of Sculptor and Fornax that are more
massive \citep{McConnachie12}; on the other hand, Sextans and Carina
have stellar masses and absolute luminosity quite similar to those of
Ursa Minor.

For this reason, in Fig. \ref{fig4} we include also 
the stars measured for these two dSph galaxies in
\citet{North12} and \citet{TJH10}; 13 more data points in total. 
Moreover, eight of these data points sit at [Fe/H]$>-1.8$,
 significantly increasing the only four data points available for Ursa Minor in
this metallicity range, where the new Herts model predicts a spread in [Mn/Fe].
Although the data belong to two different dSphs, it is still
encouraging to see that the model is in excellent agreement with them;
in fact the data for Sextans, for example, show a remarkable spread
 of more than 0.5 dex.

Therefore, we encourage more investigations to establish the presence
 of this spread in the [Mn/Fe] versus [Fe/H] space in other satellites of
 the Milky Way, in particular in the faint classical dSphs such as Ursa
 Minor, Sextans, Carina and Draco.  On the other hand, we underline
 that we cannot use the faintest satellite galaxies (or
 ultra faint galaxies) to confirm this scenario.  In fact, it is not
 clear whether all of these objects evolve enough in time to see the
 impact in their stellar population of the chemical enrichment by SNe
 Ia.

\section{Conclusions}

We present new results for the chemical evolution of the [Mn/Fe] in
a relatively faint dSph galaxy, Ursa 
Minor, with two different
prescriptions for  SNe Ia. These two prescriptions, that we call
Herts and Trieste, present the following differences: In the Trieste model, we
allow only a single channel of SNe Ia to explode, the single
degenerate with a deflagration; in the Herts model we have two
different channels, one is a sub-Ch channel, with a double
detonation, the other is a special case of a single
degenerate, originating from a relatively massive primary star, producing a
relatively weak deflagration (SN Iax channel).  These two channels produce, on
average, almost the same amount of Mn as the the single channel in the Trieste
model.  We show that in the framework of a homogenous chemical
evolution model, both Herts and Trieste prescriptions are compatible
with the data available for Mn in this dSph.  On the other hand, in
the stochastic framework, the results are different and the data
favour the Herts model, and therefore, the presence of two
channels for SNe Ia at low metallicity, in addition to normal SNe Ia
at higher metallicities.  This conclusion is supported
also by the chemical evolution results obtained for the solar vicinity
case in \citet{Seitenzahl13}.  Including the data for another two faint
dSphs with similar characteristics to Ursa Minor, Sextan and
Carina, confirms the necessity of these two channels to explain the
data, although specific chemical evolution models for these galaxies
would be best suited to check this. We plan to extend our studies also
to these galaxies as well as more massive dSphs like Sculptor to confirm
this result.  Clearly, these results need to be confirmed by future
  abundance measurements of manganese in stars of other satellite
  galaxies of our Milky Way. It will be vital for this project to
  measure not the most extreme metal-poor tail, as more commonly
  happens, but the opposite; the metal-rich end of dSph galaxies.

\begin{acknowledgements}
  This work was supported by the Science and Technology Facilities
  Council ST/M000958/1 for the BRIDGCE consortium grant.
  G.C. acknowledges financial support from the European Union Horizon
  2020 research and innovation programme under the Marie Sk\l
  odowska-Curie grant agreement No. 664931.

\end{acknowledgements}

\bibliographystyle{aa}
\bibliography{spectro}

\end{document}